\documentclass[12pt]{article}

\def\lromn#1{\uppercase\expandafter{\romannumeral#1}}

\def\blist{\begin{list}{\setlength{\rightmargin}{\leftmargin}}}
\def\elist{\end{list}}
\begin{document}

\begin{flushright}
TU/00/596 \\
KEK-TH-704\\
\end{flushright}

\begin{center}
\begin{large}

\textbf{
Quantum tunneling in thermal medium
}

\end{large}

\vspace{36pt}

\begin{large}
Sh. Matsumoto$^{1}$ and M. Yoshimura$^{2}$

$^{1}$ 
Theory Group, KEK\\
Oho 1-1 Tsukuba Ibaraki 305-0801 Japan

$^{2}$ 
Department of Physics, Tohoku University\\
Sendai 980-8578 Japan\\
\end{large}

\vspace{4cm}

{\bf ABSTRACT}
\end{center}

Time evolution of tunneling phenomena in medium is studied using
a standard model of environment interaction.
A semiclassical formula valid at low, but finite temperatures
is derived in the form of integral transform for the reduced
Wigner function, and the tunneling
probability in thermal medium is calculated for a general tunneling potential
of one dimensional system.
Effect of dissipation, its time evolution in particular, 
depends on the behavior of the potential far beyond the barrier.

\newpage

Tunneling phenomena are genuinely a quantum effect.
When they occur in some surrounding medium,
the important question arises as to whether the tunneling
rate is enhanced or suppressed by the environment effect.
There are already many works on this subject
\cite{q-tunneling review1}, and
most past works deal with a system in equilibrium as a whole.
The Euclidean technique such as the bounce solution
\cite{bounce} is often
used in this context \cite{caldeira-leggett 83},
\cite{qbm review}.
Our approach here is different, and we attempt
to clarify dynamics of time evolution starting from
an arbitrary initial state of the tunneling system,
which can be either a pure or a mixed state.
Moreover we find it more illuminating to use a real-time
formalism instead of the Euclidean method much 
employed in the literature \cite{fiss-92}.

Although one can think of many applications of our general framework
here, application to cosmology is our prime target
\cite{ew-bgeneration-review}.
The cosmological environment is in many aspects the simplest
since the environment is almost spatially homogeneous and for many purposes
its state can be characterized by a single temperature parameter.
Furthermore cosmic evolution gives rise to decrease of
the temperature, dropping some degree of freedom out of equilibrium
with the rest. This gives a natural setting for separation
of a subsystem from the environment, which is at the heart of our approach.

The standard model \cite{feynman-vernon},\cite{caldeira-leggett 83}
of environment and its interaction with
a one dimensional system which we describe by a potential $V(q)$ introduces
an infinitely many, continuous harmonic oscillators of
environment and a bilinear interaction,
\( \:
q \int d\omega \,c(\omega )Q(\omega ) \,.
\: \)
Here $Q(\omega )$ is the environment oscillator coordinate of frequency
$\omega $ and $c(\omega )$ gives a coupling strength of
the system-environment interaction.
Although the form of interaction seems rather limited, it
is generally believed that details of the environment 
interaction with the system should not be important
and this form should suffice.
With the total system specified, dynamics is given by
the quantum equation of motion,
\begin{eqnarray}
&&
\frac{d^{2}q}{dt^{2}} + \frac{dV}{dq} =
-\,\int_{\omega_c}^{\infty}\,d\omega \,c(\omega )Q(\omega ) \,, 
\hspace{0.5cm}
\frac{d^{2}Q(\omega )}{dt^{2}} + \omega ^{2}\,Q(\omega )
= -\,c(\omega )q
 \,.
\label{eq of motion}
\end{eqnarray}

Quantum Langevin equation is derived \cite{ford-lewis-oconnell}
by eliminating the environment variable
$Q(\omega \,, t)$;
\begin{eqnarray}
&&
\frac{d^{2}q}{dt^{2}} + \frac{d V}{d q} +
2\,\int_{0}^{t}\,ds\,\alpha _{I}(t - s)q(s) = F_{Q}(t) 
\,,
\label{langevin eq}
\end{eqnarray}
where $F_{Q}(t) $ has an explicit, linear dependence on the initial
values, $Q_i(\omega) \,, P_i(\omega) = \dot{Q}_i(\omega) $.
In a thermal bath of environment temperature $T = 1/\beta $
the random force from the environment is characterized by
the correlation function,
\begin{eqnarray}
&&
\langle \{ F_{Q}(\tau )\,,F_{Q}(s) \}_{+} \rangle_{{\rm env}}
= \int_{\omega _{c}}^{\infty }\,d\omega \,r(\omega )
\cos \omega (\tau - s)\,\coth (\frac{\beta \omega }{2}) \,,
\end{eqnarray}
with $r(\omega ) = c^{2}(\omega )/(2\omega )$.
The kernel function $\alpha _{I}$ in eq.(\ref{langevin eq}) 
is related to the real-time thermal Green's function and is given by
\( \:
\alpha _{I}(t) = -\,\int_{\omega _{c}}^{\infty }\,d\omega \,
r(\omega )\sin (\omega t) 
 \,.
\: \)

An often used simplification is the local, Ohmic approximation
taking 
\( \:
r(\omega) = \eta\omega/\pi
\: \)
with $\omega_c = 0$,
which amounts to
\( \:
\alpha _{I}(\tau ) = \delta \omega ^{2}\delta (\tau )
+ \eta\, \delta '(\tau ) \,.
\: \)
This gives
\( \:
\delta \omega ^{2}\,q + \eta \dot{q}
\: \)
in the Langevin equation (\ref{langevin eq}).
The parameter $\delta \omega ^{2}$ is interpreted as
a potential renormalization effect or the mass renormalization in the
field theory analogy, while $\eta $ is the Ohmic friction
coefficient.
This approximation breaks down at early times \cite{difficulty of ohm}, 
but it is useful in many other cases.

Our discussion starts from the master equation for this system,
which is written for convenience in terms of the Wigner function
$f_{W}(q \,, p\,, Q(\omega ) \,, P(\omega ))$,
a Fourier transformation of the density matrix 
\( \:
\rho (q \,, q' \,, Q(\omega ) \,, Q'(\omega ))
\: \)
with respect to the relative coordinate,
$\; q - q' \,, \; Q(\omega ) - Q'(\omega )$,
\begin{eqnarray}
\frac{\partial f_{W}}{\partial t} &=&
-\,p\,\frac{\partial f_{W}}{\partial q}
- \int\,d\omega \,
\left( \,
P(\omega )\frac{\partial f_{W}}{\partial Q(\omega )}
+ c(\omega )\,(q\frac{\partial }{\partial P(\omega )}
+ Q(\omega )\frac{\partial }{\partial p})\,f_{W}
\,\right)
\nonumber \\ 
\hspace*{0.5cm} 
&+& \frac{1}{i\hbar }\,
\left\{ \,
V\left( q + \frac{i\hbar }{2}\frac{\partial }{\partial p}\right) 
- V\left( q - \frac{i\hbar }{2}\frac{\partial }{\partial p}\right) 
\,\right\}
f_{W}
\,. \label{master eq for wigner} 
\end{eqnarray}

The crucial observation for subsequent development is
how the Planck constant $\hbar $ enters in the master 
equation (\ref{master eq for wigner}).
In one dimensional quantum mechanics the semiclassical approximation
is excellent when the potential barrier is large, and
we assume that this is also true in the presence of the 
system-environment interaction.
In the semiclassical $\hbar \rightarrow 0$ limit we have
\begin{eqnarray}
\frac{1}{i\hbar }\,
\left\{ \,
V\left( q + \frac{i\hbar }{2}\frac{\partial }{\partial p}\right) 
- V\left( q - \frac{i\hbar }{2}\frac{\partial }{\partial p}\right) 
\,\right\}
f_{W}
\,\rightarrow\,
\frac{d V}{d q}\,\frac{\partial f_{W}}{\partial p}
\,.
\end{eqnarray}
The resulting equation, being identical to the classical Liouville equation,
has an obvious solution;
\begin{eqnarray}
&&
f_{W}(q\,, p\,, Q\,, P) = 
\int\,dq_{i}dp_{i}\,\int\,dQ_{i}dP_{i}\,
f_{W}^{(i)}(q_{i}\,, p_{i}\,, Q_{i}\,, P_{i})\,
\nonumber \\ &&
\hspace*{1cm} 
\cdot 
\delta \left( q - q_{{\rm cl}}\right)\,\delta \left( p - p_{{\rm cl}}\right)
\,\delta \left( Q - Q_{{\rm cl}}\right)\,\delta \left( P 
- P_{{\rm cl}}\right)
 \,,
\end{eqnarray}
where 
\( \:
q_{{\rm cl}}(q_{i}\,, p_{i}\,, Q_{i}\,, P_{i}\, ; t)
\: \)
etc. are the solution of 
(\ref{eq of motion}), taken as the classical equation.

We consider the circumstance under which the tunneling system
is initially in a state uncorrelated to the rest of environment.
Thus we take an uncorrelated initial state of the form
\( \:
\rho ^{(i)} = \rho _{q}^{(i)} \otimes \rho _{Q}^{(i)}
 \,,
\: \)
to get the reduced Wigner function after the trivial
$Q(\omega) \,, P(\omega)$ integration,
\begin{eqnarray}
&&
f_{W}^{(R)}(q \,, p\,;t) = 
\int\,dq_{i}dp_{i}\,
f_{W\,, q}^{(i)}(q_{i}\,, p_{i})\,K(q \,, p\,, q_{i}\,, p_{i}\,;t)
\,,
\nonumber \\ && 
K(q \,, p\,, q_{i}\,, p_{i}\,;t) = \int\,dQ_{i}dP_{i}\,
f_{W\,, Q}^{(i)}(Q_{i}\,, P_{i})\,
\delta \left( q - q_{{\rm cl}}\right)\,\delta \left( p - p_{{\rm cl}}\right)
 \,.
\end{eqnarray}
The problem of great interest is how further one can simplify
the kernel function $K$ here.

In many situations one is interested in the tunneling probability when
the environment temperature is low enough.
At low temperatures of
\( \:
T \ll 
\: \)
a typical frequency or curvature scale 
$\omega _{s}$ of the potential $V$,
one has
\( \:
\omega _{s}\,\sqrt{\,\overline{Q_{i}^{2}(\omega )}\,} \,, 
\sqrt{\,\overline{P_{i}^{2}(\omega )}\,} = O[\sqrt{T}]
\ll \sqrt{\omega _{s}}
 \,.
\: \)
Expansion in terms of 
\( \:
Q_{i}(\omega ) \,, P_{i}(\omega )
\: \)
is then justified.
Thus, we use
\begin{eqnarray}
&&
\delta \left( \,q - q_{{\rm cl}}\,\right)
= \int\,\frac{d\lambda _{q}}{2\pi }\,
\exp \left[ \,i\,\lambda _{q}\,\left( \,q - q_{{\rm cl}}\,\right)\,\right]
\nonumber \\ &&
\hspace*{-1cm}
\approx \int\,\frac{d\lambda _{q}}{2\pi }\,
\exp \left[ \,i\,\lambda _{q}\,
\left( \,q - q_{{\rm cl}}^{(0)} -
\int\,d\omega \,
\left\{\,Q_{i}(\omega )q_{{\rm cl}}^{(Q)}(\omega )
+ P_{i}(\omega )\,q_{{\rm cl}}^{(P)}(\omega )\,
\right\}\,\right)\,\right]
 \,,
\label{expansion of cl sol}
\end{eqnarray}
valid to the first order of $Q_{i}(\omega) \,, P_{i}(\omega)$.
A similar expansion for
\( \:
\delta \left( \,p - p_{{\rm cl}}\,\right)
\: \)
using
\( \:
p_{{\rm cl}}^{(0)} \,, p_{{\rm cl}}^{(Q)}(\omega ) \,, 
p_{{\rm cl}}^{(P)}(\omega )
 \,,
\: \)
also holds.

Gaussian integral for the variables
\( \:
Q_{i}(\omega ) \,, P_{i}(\omega ) \,, \lambda _{q} \,, \lambda _{p}
\: \)
can be done explicitly with eq.(\ref{expansion of cl sol}), 
when one takes the thermal density matrix
for the initial environment, $\rho _{Q}^{(i)}$.
The result of this Gaussian integral leads to
an integral transform of the Wigner function, 
$f_{W}^{(i)} \rightarrow f_{W}^{(R)}$, using the kernel function of
\begin{eqnarray}
&&
\hspace*{-0.5cm}
K(q \,, p\,, q_{i}\,, p_{i}\,;t) = 
\frac{\sqrt{{\rm det}\; {\cal J}\,}}{2\pi}
\exp \left[ \,-\,\frac{1}{2}
(q - q_{{\rm cl}}^{(0)}\,, \, p - p_{{\rm cl}}^{(0)})
\,{\cal J}\,
\left( \begin{array}{c}
q - q_{{\rm cl}}^{(0)}  \\
p - p_{{\rm cl}}^{(0)}
\end{array}
\right)
\,\right]
 \,,
\end{eqnarray}
where the matrix elements of 
\( \:
{\cal J}^{-1} = I_{ij}
\: \)
are given by
\begin{eqnarray}
&&
I_{11} =\frac{1}{2}\,\int_{\omega _{c}}^{\infty }\,d\omega \,
\coth \frac{\beta \omega }{2}\,
\frac{1}{\omega } \,|z(\omega \,, t)|^2
\,, 
\\ &&
I_{22} =\frac{1}{2}\,\int_{\omega _{c}}^{\infty }\,d\omega \,
\coth \frac{\beta \omega }{2}\,
\frac{1}{\omega } \,
|\dot{z}(\omega \,, t)|^2
\,, 
\\ &&
\hspace*{-1cm}
I_{12}  =\frac{1}{2}\,\int_{\omega _{c}}^{\infty }\,d\omega \,
\coth \frac{\beta \omega }{2}\,
\frac{1}{\omega } \,\Re\left( z(\omega \,, t)\dot{z}^*(\omega \,, t)\right)
\,,
\end{eqnarray}
where
\( \:
z(\omega \,, t) = 
q_{{\rm cl}}^{(Q)}(\omega \,, t) + i\,
\omega \,q_{{\rm cl}}^{(P)}(\omega \,, t)\,,
\: \)
and 
\( \:
\dot{z}(\omega \,, t) = 
p_{{\rm cl}}^{(Q)}(\omega \,, t) +
i\omega \,p_{{\rm cl}}^{(P)}(\omega \,, t)
\,.
\: \)

Quantities that appear here ought to be determined by solving
differential equations with a given initial condition; 
the homogeneous Langevin equation and its linearized equation of the form 
\begin{eqnarray}
&&
\frac{d^{2}q_{{\rm cl}}^{(0)}}{dt^{2}} +
\left( \frac{d V}{dx }\right)_{q_{{\rm cl}}^{(0)}}
+ 2\,\int_{0}^{t}\,ds\,\alpha _{I}(t - s)q_{{\rm cl}}^{(0)}(s) = 0
 \,,
\\ &&
\hspace*{-1cm}
\frac{d^{2}z(\omega \,, t)}{dt^{2}} +
\left( \frac{d^{2} V}{dx^{2} }\right)_{q_{{\rm cl}}^{(0)}}\,z(\omega \,, t)
+ 2\,\int_{0}^{t}\,ds\,\alpha _{I}(t - s)z(\omega \,, s) = 
-\,c(\omega )e^{i\omega t}
\,.
\end{eqnarray}
The same linear equation as for $z(\omega \,, t)$ holds for
\( \:
\dot{z}(\omega \,, t)  \,.
\: \)
The initial condition is \\
\( \:
q_{{\rm cl}}^{(0)}(t = 0) = q_{i} \,, \hspace{0.5cm} 
p_{{\rm cl}}^{(0)}(t = 0) = p_{i} \,, \hspace{0.5cm} 
z(\omega \,, t = 0) = 0  \,, \hspace{0.5cm}
\dot{z}(\omega \,, t = 0) = 0 \,.
\: \)
Once the reduced Wigner function is known, one can compute any
physical quantity by a phase space integration.

The physical picture underlying the formula for integral
transform should be evident;
the probability at a phase space point $(q\,,p)$ is dominated 
by the semiclassical trajectory (environment effect of dissipation 
being included for its determination)
reaching $(q\,,p)$ from an initial point $(q_i \,, p_i)$ 
whose contribution is weighed by the quantum mechanical probability 
$f^{(i)}_W$ initially given.
The  broadning of the trajectory due to the environment
interaction is given by the width factor $\sqrt{I_{ij}}$.

To proceed further, we distinguish two cases of the potential type,
depending on the value of $V(\infty)$ relative to the local minimum value 
$V_m$ in the potential well.
The kind of potential we have in mind is depicted in Fig.1.
When $V(\infty) < V_m$ (Fig.1a), the classical motion is usually monotonic
ending at $x = \infty$,
while for $V(\infty) > V_m$ (Fig.1b) the motion is a damped oscillation
towards $x_0$ at the true minimum, unless the friction is large.
If the friction is larger than a critical value of
$\approx 2\omega_0$ with $\omega_0$ given by the curvature of the
potential at the global minimum, there occurs the overdamping 
such that $q_{\rm{cl}}^{(0)} \rightarrow x_0$ monotonically.
A typical interesting case for $V(\infty) > V_m$
is the asymmetric double well as may occur in the first order
electroweak phase transition \cite{ew-bgeneration-review}.

In the rest of this paper, we consider a few problems
to illustrate consequences of our general formula of
the integral transform.
One problem is the outgoing flux to the overbarrier region
of the type of potential of Fig.1a, assuming an initial 
energy eigenstate under the whole potential $V$. 
The other problem is the tunneling probability for the type of
potential of Fig.1b. We take here an initial thermal
state of the same temperature as the environment,
given by the density matrix
\( \:
\propto \sum_n e^{- \beta E_n}\,|n\rangle \langle n|
\: \)
with $|n\rangle$ the exact energy eigenstate under $V$.
This choice is made in order to mimic the cosmological first order
phase transition in which the local minimum in the left
was the true minimum above a critical temperature.
We insist on the energy eigenstate because it is the most stable
state, being stationary when the environment interaction is
switched off. In this sense the choice of the energy eigenstate
and its mixture is useful in many applications.

We first discuss the tunneling probability for the case of
$V(\infty) < V_m$. 
Suppose that we are primarily interested in the
tunneling probability from a state localized in the left well
into the right region far beyond the barrier.
For discussion from further on we assume $\omega_* \ll V_h$,
where $\omega_*$ is the frequency defined at the potential bottom in
the well and $V_h$ the barrier height seen from the bottom.
Since higher energy states contribute roughly with the weight factor,
\( \:
e^{- 2H(q_i \,, p_i)/\omega_*}\,,
\: \)
where $H(q_i \,, \, p_i)$ is the hamiltonian of the hypothetical oscillator
of frequency $\omega_*$ without the barrier in the right,
the high momentum component above the barrier is suppressed by
$e^{- \,2 V_h/\omega_*}$.
We assume that this factor is much smaller than a typical barrier penetration 
factor $|T(E)|^2$ in the overbarrier region, usually given by
the WKB formula,
\( \:
\exp[- 2\int\,dx |p(x)|\,] \,.
\: \)
In this way we may restrict the phase space region in the integral
transform to the region $D$ defined by
\( \:
|V_0| < 
\frac{1}{2}\,p_i^2 + \frac{1}{2}\,\omega_0^2\,(q_i - x_0)^2 
< V_h + |V_0| \,,
\: \)
with $V_0$ the minimum potential value at $x_0$.

The flux at $x$ is calculated from an integral of the form,
\( \:
\sim \int\,dp\,
p\,f_{W}^{(R)}(x\,, p\,; t)
\,,
\: \)
giving
\begin{eqnarray}
I(x\,, t) &=& 
\int_D\,dq_{i}\,dp_i\,f_W^{(i)}(q_i \,, p_i)\,
\frac{1}{\sqrt{2\pi I_{11}}}\,\exp \left[ \,-\,
\frac{(x - q_{{\rm cl}}^{(0)})^{2}}{2I_{11}}\,\right]
\nonumber \\ 
&&
\cdot \left( \,p_{{\rm cl}}^{(0)} + \frac{I_{12}}{I_{11}}\,
(x - q_{{\rm cl}}^{(0)})\,\right)
 \,.
\label{flux formula}
\end{eqnarray}
The initial Wigner function 
$f_{W}^{(i)}(q_{i} \,, p_{i})$ is a Fourier transform of
\( \:
\psi^{*}(q_{i} - \frac{\xi }{2})\psi(q_{i} + \frac{\xi }{2})
\: \)
with respect to $\xi$, where $\psi$ is the initial wave function.
The phase of the wave function plays an important role for the
region, $q_i > x_*$ where $x_*$ is a turning point in the overbarrier
region, and
one may use the stationary phase approximation for the $\xi$ and 
$p_i$ integration. We find a stationary point at
\( \:
\xi = 0 \,, \hspace{0.5cm} 
p_{i} = I(q_{i}) \,, \hspace{0.3cm} 
I(x) =
\frac{-\,i}{2}\,\left(\,\psi^{*} (x)\psi'(x) -
\psi'\,^{*}(x)\psi (x)\,\right)/|\psi (x)|^{2}
 \,.
\: \)
The quantity $I(x)$ is the usual quantum mechanical flux
associated with the initial quantum state.

In the region of $ x \gg x_*$ there is always a classical trajectory that 
reaches the point $x$ of the Gaussian peak from 
an initial $q_{i}$ in the range 
$q_i > x_{*}$, and the entire region within the Gaussian width 
$\sqrt{I_{11}}$ is fully covered by the integral.
The factor outside the Gaussian exponent is well approximated
by its peak value in the weak coupling case.
Thus, one obtains
\( \:
I(x\,, t) \approx |\psi (x_{c})|^{2}\,\left( -\,\dot{x}_{c}\right)
\,,
\: \)
where one determines $x_{c}(x\,, t) $ from
\( \:
q_{{\rm cl}}^{(0)}(x_{c} \,, I(x_{c}) \,, 0 \,, 0 \,; t)
= x \,.
\: \)
Using the WKB wave function for the energy eigenstate
at $q_i > x_*$,
\( \:
\psi (q_{i}) = T(E)\exp \left[ i\,\int_{x_{*}}^{q_{i}}\,p(x)\right]
/\sqrt{p(q_{i})}
 \,, \;
p(x) = \sqrt{\,2(E - V(x))\,}
\,,
\: \)
one has
a factorized form of the flux;
\begin{eqnarray}
&&
I(x\,, t) \approx |T(E)|^{2}\,f(x\,, t\,; E) \,, 
\hspace{0.5cm} 
f = \frac{p_{{\rm cl}}^{(0)}(x_{c} \,, t)}{p(x_{c})}
\left( \frac{dq_{{\rm cl}}^{(0)}(x_{c} \,, t)}{dx_{c}}
\right)^{-1}
 \,.
\end{eqnarray}
We note that the potential renormalization effect should be included 
for $|T(E)|^{2}$, as emphasized in \cite{caldeira-leggett 83}.

This formula for the flux reproduces our previous result for a specific
potential of the inverted harmonic oscillator \cite{my-00-1},
\( \:
V(x) = -\frac{1}{2}\,\omega_R ^2\,x^2 \,.
\: \)
In this case the Gaussian integral gives the exact result,
relying neither on the low temperature nor on the stationary
phase approximation. The present approach actually improves 
our previous result;
\begin{eqnarray}
&&
f = \frac{\stackrel{..}{g}x_{c} + \dot{g}I(x_c)}
{\dot{g}I(x_c) + g\omega _{B}^{2}x_{c}}
\rightarrow \frac{\stackrel{..}{g} + \omega _{B}\dot{g}}
{\omega _{B}(\dot{g} + \omega _{B}g)}
 \,,
\end{eqnarray}
where $\omega_B \approx \omega_R  - \eta/2$ is a diagonalized frequency,
and the initial flux \\
$I(x_c) = \sqrt{2(E - V(x_{c}))}$ in the WKB approximation.
The function $g(t)$ is the homogeneous solution of
the Langevin equation given in an explicit form in \cite{my-00-1}.
The limiting formula is valid in the infinite $x$ limit
as derived in \cite{my-00-1}.
Both at early and late times the factor $f \approx 1$,
deviating from unity only for the time range of order $1/\omega_B$.

For discussion of a more general case of finite $V(\infty) < V_m$ 
we use the local, Ohmic approximation, 
which becomes excellent at late times.
A potential that decreases fast at infinity
as $x \rightarrow \infty $ is assumed;
\( \:
\frac{dV}{dx} \rightarrow 0 \,.
\: \)
The acceleration term can then be neglected if
$|\frac{d^{2}V}{dq_{{\rm cl}}^{2}}| \ll \eta ^{2}$.
This is a slow rolling approximation, and it always holds
for $x$ large enough.
The classical equation is then solved as
\( \:
\eta \,\int_{x_{c}}^{x}\,dz(\frac{dV}{dz})^{-1} = - \,t
 \,,
\: \)
which gives the factor
\( \:
f \approx 
-\,(\frac{dV}{dx_{c}})/(\eta p(\infty ))  \,.
\: \)
Thus, the tunneling probability decreases with time along with the decreasing
slope of the potential.
Since the tunneling may not be terminated within a limited finite
time, this result poses a curious question;
there is a situation in the early universe
in which the tunneling is never ended if the friction is strong.

We next turn to the case of $V(\infty) > V_m$,
and consider the thermal initial state.
At low temperatures of $T \ll V_h$ the dominant contribution
comes from the energy range $0 < E < V_h$.
Contribution from the subbarrier region in $q_i$ integration
is small and may be ignored.
We add two contributions from the left well and from the overbarrier
region.
In the left well region the Wigner function is given by
\( \:
f_W ^{(i)}(q_i \,, p_i) \propto 
\exp[- \tanh(\beta \omega_*/2)(\omega_* q_i^2 +
p_i^2/\omega_*)]
\,.
\: \)
It is more appropriate here to compute the transition probability by
\( \:
\sim \int\,dp\,
f_{W}^{(R)}(x \,, p\,; t)
\,,
\: \)
without the flux $p$ factor.
The final position $x$ is taken around $x_0$ of width $\Delta x$.
If one makes a rough approximation for both the left well
and the overbarrier  region by harmonic oscillators,
the relevant classical motion is given by a linear function
of initial values, $q_i \,, p_i$, reducing the phase space
integral to a restricted Gaussian type.

The averaged tunneling probability over $\Delta x$
(Gaussian weight assumed) is \\
\( \:
\frac{1}{\sqrt{\pi A}}\,\exp\left( -\,\frac{x_0^2}{A} \right)
\: \)
from the left well region, with
\( \:
A = (\frac{g_1^2}{\omega_*} + \omega_* g_2^2)\coth(\frac{\beta \omega_*}{2})
 + 2(\Delta x^2 + I_{11}) \,,
\: \)
where $g_i(t)$'s are two independent solutions to the linearized
equation in the left well.
The time averaged value over $\Delta t \gg 1/\omega_*$ gives
$\overline{g_1^2} \approx \frac{1}{2}e^{-\eta t}$, and
$\overline{g_2^2} \approx e^{-\eta t}/(2\omega_*^2) $,
in the weak coupling.
In the late time limit 
\( \:
I_{11} \approx \coth(\beta \omega_*/2)/(2 \omega_*)
\,,
\: \)
and the exponent factor becomes of order,
\( \:
\exp[- \omega_* x_0^2/(1 + 2\omega_* \Delta x^2)]
\: \)
at $T \ll \omega_*$, while at $T \gg \omega_*$ it is
\( \:
\exp[- \omega_*^2 x_0^2/(2 T + 2\omega_*^2 \Delta x^2)]
\,.
\: \)

On the other hand, one may use
the WKB form of the wave function in the overbarrier region,
with the transmission coefficient
\( \:
|T(E)|^2 = e^{-\,2\pi(V_h - E)/\omega_B}
\,,
\: \)
using the inverted harmonic oscillator of the curvature
$\omega_B$ for the subbarrier region.
The density matrix $\rho_q^{(i)}(q \,,q')$ is then
($T \gg \omega_*$ is assumed)
\begin{eqnarray}
&&
\beta \omega_* e^{- 2\pi V_h/\omega_B}\,\int_{0}^{V_h}\,dE\,
\exp\left[-\,(\beta - \frac{2\pi}{\omega_B})\,E \right]\,
\varphi(q \,; E)\varphi^*(q' \,; E)
\,,
\end{eqnarray}
where 
\( \:
\varphi(q \,; E) \approx
\cos(\int_{x_*}^{q}dx\,p(x\,; E) - \frac{\pi}{4})/\sqrt{p(q \,;E)}
\,.
\: \)
The contribution from the overbarrier region is then roughly \\
\( \:
\beta \omega_*(e^{-2\pi V_h/\omega_B} - e^{-\beta V_h})
/(\,(\beta - 2\pi/\omega_B)
(\omega_0 \sqrt{\, 2\pi \Delta x^2 + \pi \coth(\beta \omega_0/2)/\omega_0\,})
\,)
\,.
\: \)
In the case of overdamping this asymptotic value is approached from
below by a slowly varying term 
\( \:
\propto e^{- 2\omega_0^2 t/\eta} \,.
\: \)
Since $\omega_* x_0^2 \gg V_h$ and $\omega_* = O[\omega_B]$,
these contributions from the overbarrier region are generally
larger than those from the left well region.
One should compare this with the zero temperature result,
\( \:
\omega_* e^{-2\pi V_h/\omega_B}/
(\omega_0 \sqrt{\, 2\pi \Delta x^2 + \pi/\omega_0})
\: \)
in order to assess the effect of dissipative medium.

\vspace{0.5cm}
In summary we gave a real-time formulation of tunneling
dynamics in thermal medium within the semiclassical
framework.

\vspace{1cm}
\begin{center}
{\bf Acknowledgment}
\end{center}

An important part of
this work was completed during our stay at DESY, and
both of us acknowledge the theory group of DESY, especially
W. Buchmuller for the warm hospitality.
The work of Sh. Matsumoto is partially
supported by the Japan Society of the Promotion of Science.

\vspace{1cm} 

\newpage
\begin{Large}
\begin{center}
{\bf Figure caption}
\end{center}
\end{Large}

\vspace{0.5cm} 
\hspace*{-0.5cm}
{\bf Fig.1}

Two types of tunneling potential.

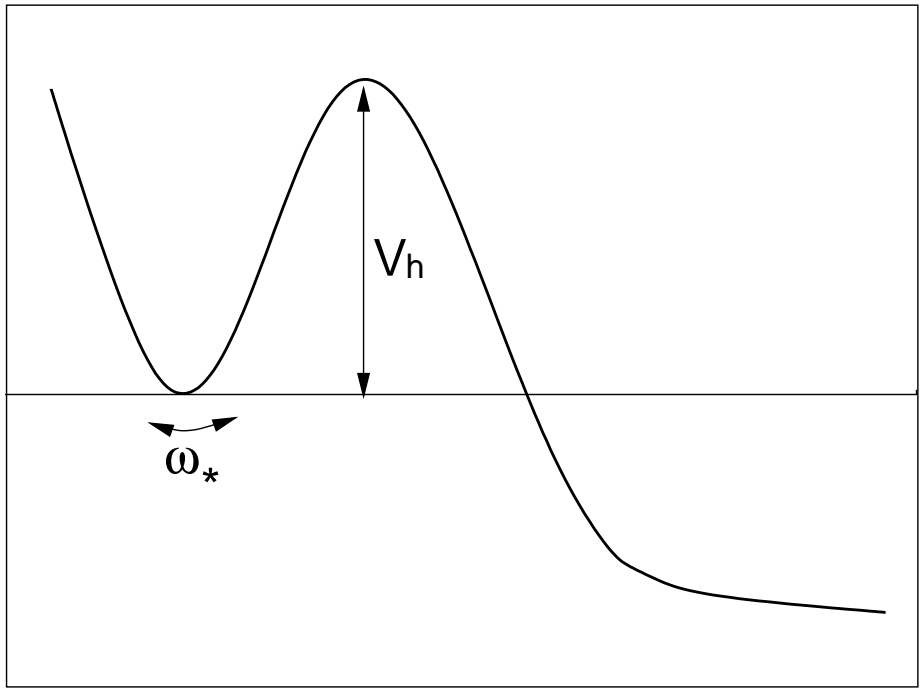
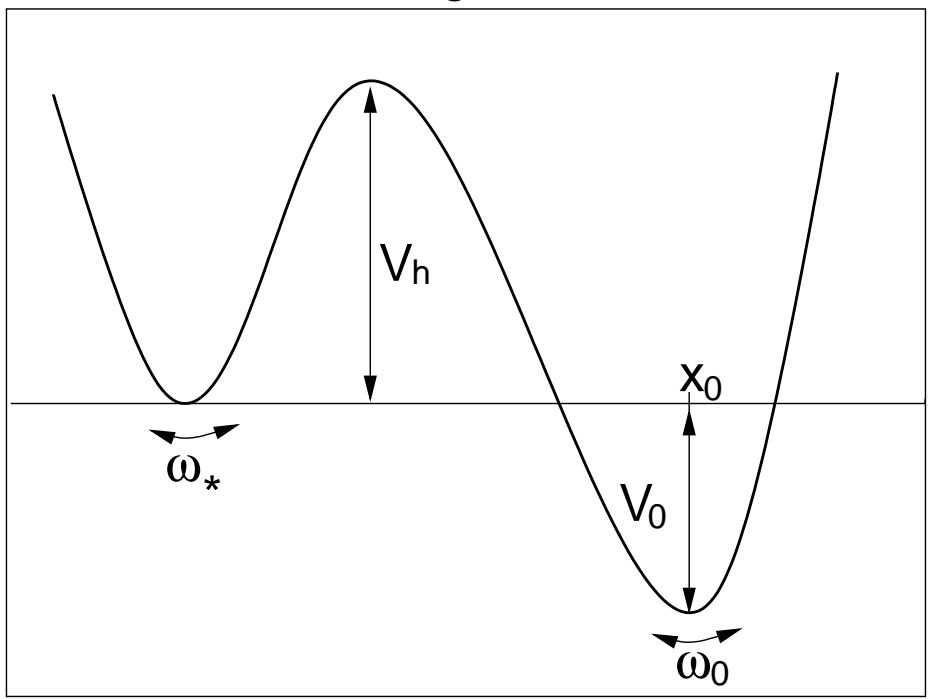
\end{document}